# All-Electrical Control of a Hybrid Electron Spin/Valley Quantum Bit in SOI CMOS Technology

L. Hutin[1*], L. Bourdet[2], B. Bertrand[1], A. Corna[2], H. Bohuslavskyi[1,2], A. Amisse[1,2], A. Crippa[2], R. Maurand[2], S. Barraud[1], M. Urdampilleta[3], C. Bäuerle[3], T. Meunier[3], M. Sanquer[2], X. Jehl[2], S. De Franceschi[2], Y.-M. Niquet[2*], M. Vinet[1]

[1] CEA, LETI, Minatec Campus, F-38054 Grenoble, France
[2] CEA, INAC, F-38054 Grenoble, France    [3] CNRS, Institut Néel, F-38042 Grenoble, France
*e-mail: louis.hutin@cea.fr; yann-michel.niquet@cea.fr

## Abstract
We successfully demonstrated experimentally the electrical-field mediated control of the spin of electrons confined in an SOI Quantum Dot (QD) device fabricated with a standard CMOS process flow. Furthermore, we show that the Back-Gate control in SOI devices enables switching a quantum bit (qubit) between an electrically-addressable, yet charge noise-sensitive configuration, and a protected configuration.

## Introduction

Following the emergence of Silicon spin qubits as serious contenders in the race for quantum computation [1], we have recently demonstrated two-axis control of the first hole spin qubit in Si transistor-like structures using a CMOS technology platform [2,3]. Tunnel barriers are defined by protecting the SOI film from self-aligned ion implantation between dense Gates (64nm pitch) by larger-than-usual SiN spacers (typically 30nm), thus leading to a linear arrangement of Gates along an intrinsic NanoWire [4] (**Fig. 1**). At very low temperatures (~1K and below), each Gate defines a QD with a discrete energy spectrum, which can be used to confine a small number of charges controlled by the Coulomb blockade effect (**Fig. 2**). Making a qubit out of a QD entails the ability to initialize and manipulate a two-level quantum state of a single charge, such as spin-down $|\downarrow\rangle$ and spin-up $|\uparrow\rangle$.

Inducing Electron Spin Resonance (ESR) with an RF magnetic-field is the most straightforward approach to spin control (**Fig. 3**), although the excitation is hardly applied locally, as opposed to an all-electrical control scheme. The latter however requires a way to couple the spin of a charge to its orbital motion. Unfortunately in Si electrons, unlike holes, have generally weak intrinsic Spin-Orbit Coupling (SOC). Electrical Dipole Spin Resonance (EDSR) can in principle be achieved by placing the charge in a magnetic-field gradient produced by a micromagnet [5], though more compact and scalable alternatives are desirable.

## Device and definition of quantum states

Our test device for EDSR demonstration (**Fig. 4**) consists of a Two-Gate nFET-like structure with Gates partially wrapping around the [110]-oriented SOI NanoWire (W=30nm; $T_{Si}$=12nm). We consider two QDs, QD1 and QD2 confined in the "corners" defined by $G_1$ and $G_2$. If both are in the same spin state (*e.g.* parallel spins, which is the ground state in a finite magnetic field B), Pauli's exclusion principle prevents charge movement from QD1 to QD2, and hence $I_{DS}$ current from flowing. However, a spin rotation obtained by applying a resonant RF E-Field to $G_1$ would lift the Pauli Spin Blockade and enable a non-zero current. Spin degeneracy is lifted by means of an externally-applied static magnetic field, the splitting energy being $E_Z=g.\mu_B.B$ where g is the Landé g-factor (g≈2 for electrons in Si) and $\mu_B$ the Bohr magneton. The principle of resonant spin transitions, and the corresponding expected ESR signal are shown on **Fig. 5**. Yet, the additional valley degree of freedom needs to be considered. The conduction band of bulk Si features six degenerate $\Delta$ valleys. Structural and electrical confinement in our device, however, leaves two low-lying valleys $v_1$ and $v_2$, projected in $\Gamma$ and separated by an energy $\Delta_V$ (**Fig. 6**). From these two valleys, four distinct states can be resolved upon applying a static magnetic field: $|v_1,\downarrow\rangle, |v_1,\uparrow\rangle, |v_2,\downarrow\rangle$ and $|v_2,\uparrow\rangle$.

## Corner Dots and spin-valley mixing

Of particular interest are the two states $|v_1,\uparrow\rangle$ and $|v_2,\downarrow\rangle$, which may be mixed under the condition that an inter-valley spin-orbit (SO) coupling $C_{v1v2}$ in the Hamiltonian is non-zero. As illustrated in **Fig. 7**, this criterion is fulfilled if the mirror symmetry of the electron wavefunction with respect to the (XZ) plane is broken. The partially overlapping Gate leading to the "Corner Dot" confinement is therefore the key to spin-valley-orbit mixing in this case.

As B is increased and the spin splitting $E_Z=g.\mu_B.B$ approaches the valley splitting $\Delta_V$, the $|v_1,\uparrow\rangle$ and $|v_2,\downarrow\rangle$ energies may either cross (no coupling) or anticross ($C_{v1v2} \neq 0$). In the former case (**Fig. 8a**), only spin-preserving inter-valley transitions can be expected in response to pure E-field excitations. In the latter case, due to states mixing near the anticrossing, B-dependent spin/valley transition diagonals may add-up to the EDSR signal (**Fig. 8b**). A color plot of $I_{DS}$ measured in a cryostat at T=15mK vs. E-field frequency and B clearly shows spin resonance lines (**Fig. 8c**). This is to our knowledge the first experimental measurement of micromagnet-free resonant E-field manipulation of electron spins in Si QDs [6].

## Programming a valley state, encoding a spin state

Since the splitting between $v_1$ and $v_2$ is related to charge confinement close to an interface, it is possible to tune $\Delta_V$ by modulating the vertical electric field. This was shown in [7] using coplanar side Gates on bulk Si, but SOI offers the possibility of using the Back-Gate potential $V_b$. We calculated the $\Delta_V(V_b)$ energy dependence using a Tight Binding model for the valley and the SO coupling at the atomistic level [8]. The results are shown in **Fig. 9** together with corresponding plots of the electron wavefunction. The tunability of $\Delta_V$ can be leveraged as schematized on **Fig. 10**: adiabatically changing $V_b$ allows following the lower branch past the anticrossing and transitioning continuously from $|v_1,\uparrow\rangle$ to $|v_2,\downarrow\rangle$. If one defines the qubit basis states $|0\rangle$ as $|v_1,\downarrow\rangle$ and $|1\rangle$ as this hybridized lower branch, $V_b$ enables to switch between a pure spin regime and a pure valley regime. The advantage of a valley qubit is the all-electrical addressability of inter-valley transitions, the downside being sensitivity to charge noise and hence shorter decoherence times. Conversely, when in spin regime, the qubit is scarcely addressable electrically but benefits from a longer lifetime.

This approach leads to circumventing a trade-off between qubit manipulation speed and coherence time, thus improving the number of operations/error. Advantageously, the qubit rotation speed is maximal when the charge is pulled away from the interfaces, which is more difficult to achieve by using only coplanar Front Gates [9]. **Fig. 11** shows the simulated chronograms of the electrical RF Gate1 excitation signal ($\nu$ = 23.66 GHz), the resulting Rabi oscillations of the qubit ($f_{Rabi}$ = 80 MHz) in valley mode, and the eventual spin rotation as $V_b$ adiabatically ramps past the anticrossing back to spin mode. We accounted for local surface roughness variability to estimate the impact of $\Delta_V$ fluctuations on the optimal operating $V_b$ range (**Fig. 12**), which can be individually calibrated for each qubit with separate Back-Gates.

## Conclusions

We induced spin transitions in MOS Gate-confined electrons in a Si NW using only E-field excitations and without resorting to co-integrated micromagnets. The underlying mechanism is based on the interplay between Spin-Orbit Coupling (SOC) and the multi-valley structure of the Si conduction band, and is enhanced by the "Corner Dot" device geometry. By offering the ability to break and restore the confinement symmetry at will, the SOI Back-Gate permits fast programming in valley mode, and information storage in spin mode. This functionality could alleviate the trade-off between fast manipulation and long coherence time, thereby improving the outlook for compact, scalable and fault-tolerant quantum logic circuits.

## Acknowledgment

The authors acknowledge financial support from the EU under Project MOS-QUITO (No. 688539).

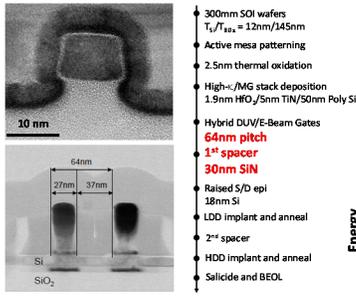
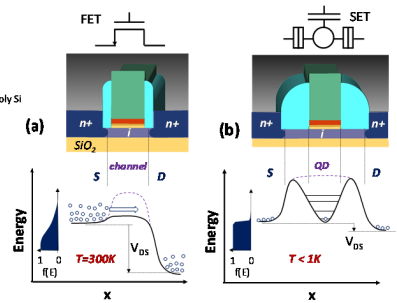
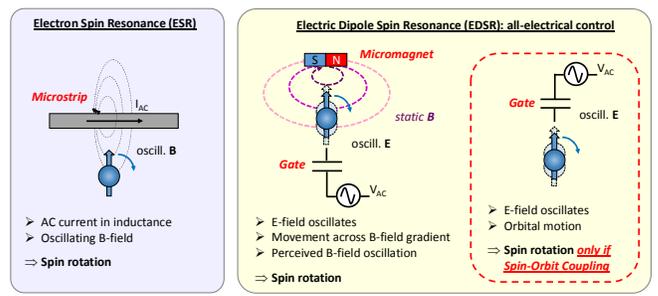

Fig.1: Top left: STEM view along the Gate wrapping around the Si channel. Bottom left: STEM view of two Gates in series (64nm pitch) showing the width of the 1st spacer. Right: simplified process flow.

Fig.2: Energy profile along the channel of (a) an SOI FET at 300K in which carriers flow continuously above a lowered barrier (b) an SET operating in the Coulomb Blockade regime at low T due to large tunnel barriers beneath the spacers and energy quantization in the Gate-defined Quantum Dot (QD).

Fig.3: Schematic description of various methods to induce spin transitions for a localized charge. Magnetic field manipulation is physically straightforward but requires flowing an AC current through a microstrip in the vicinity of the targeted spin. Creating a B-field gradient through a micromagnet enables indirect control via electric field. A more scalable, all-electrical micromagnet-free approach is possible in the case of strong spin-orbit coupling, which in Si usually applies to holes but not electrons.

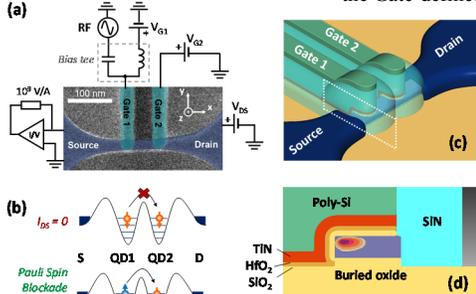
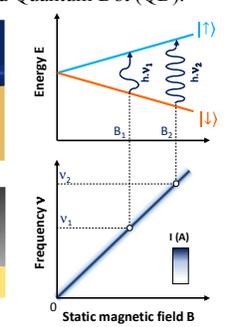
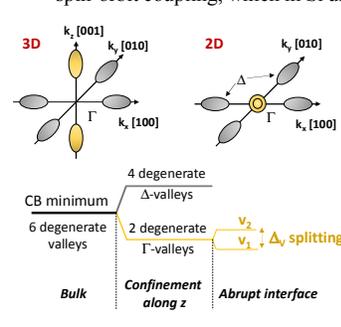
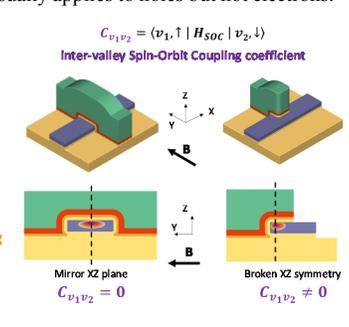

Fig.4: (a) Top-view SEM of the two-Gate device after Gate patterning, and setup description. (b) Spin-filtering mechanism across the Double QD based on the Pauli Spin Blockade rectifying the Drain current. (c) Schematic view of the partially wrapping Gates. (d) Cross-section along a Gate and representation of the asymmetrical electron wavefunction along the mesa edge, or "Corner Dot".

Fig.5: Principle of Zeeman splitting of the degenerate spin states, and of resonant transitions. Bottom shows a typical E(D)SR signal: current is prevented by the Pauli Spin Blockade except on the transition line.

Fig.6: Valley splitting in a 2D-confined configuration in Silicon. The originally sixfold degenerate Δ valleys split into four Δ in the plane of confinement, and two Γ. The abrupt interface further splits the two Γ valleys into $v_1$ and $v_2$ by an energy noted $\Delta_V$ in the following.

Fig.7: Impact of device geometry on inter-valley Spin-Orbit Coupling. The coupling term $C_{v1v2}$ is non-zero if the symmetry of the electron wavefunction is broken the (XZ) plane. This condition is fulfilled in the case of Corner Dots.

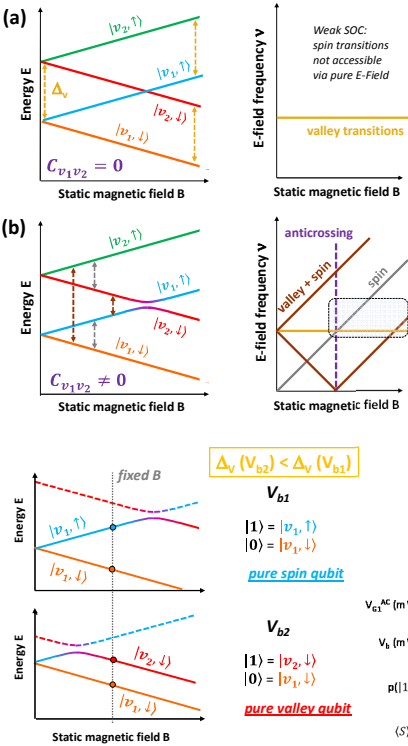
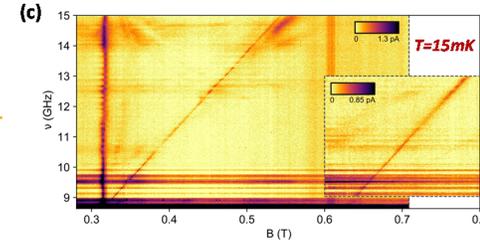
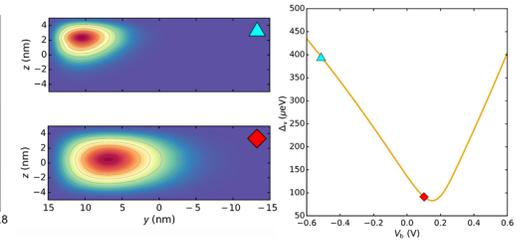

Fig.8: (a) Zeeman splitting from $v_1$ and $v_2$ in the case of no inter-valley SOC, and associated expected EDSR signal. (b) Case in which inter-valley SOC exists and states anti-cross, and expected EDSR. The dotted frame symbolizes the measured region. (c) Experimentally measured EDSR signal, showing spin and spin/valley transitions.

Fig.9: Simulated influence of the SOI Back-Gate voltage $V_b$ on the $\Delta_V$ valley splitting for an ideal device (no surface roughness). $\Delta_V$ is maximal when the charge is confined against an interface. Positive $V_b$ tends to pull the wavefunction towards the center of the NanoWire, away from the interfaces. A further $V_b$ increase results in increasing $\Delta_V$ again, due to charge confinement against the interface with the buried oxide.

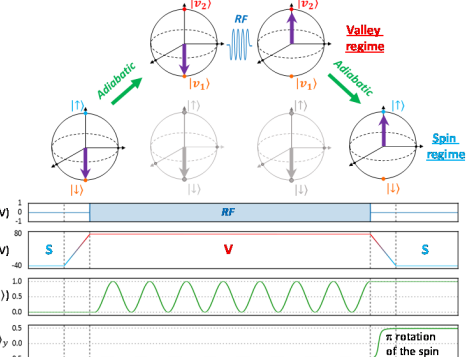
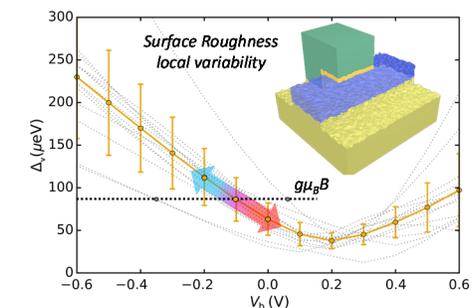

Fig.10: Energy diagram showing two $V_b$ configurations. At a given B, changing $V_b$ adiabatically enables to switch between a spin qubit and a valley qubit regime, by operating respectively left and right of the anti-crossing.

Fig.11: Simulated purely electrical manipulation of the spin of a confined electron. A $V_b$ ramp brings the qubit in the valley regime, in which it can oscillate ($f_{Rabi}$ = 80MHz) in response to an RF E-field excitation (here ν=23.66 GHz). As the $V_b$ ramp is reversed, the $|1\rangle$ eigenstate transitions from $|v_2,\downarrow\rangle$ to $|v_1,\uparrow\rangle$, thus leading to a π rotation of the spin.

Fig.12: Impact of local surface roughness variability on the $\Delta_V(V_b)$ dependence (RMS 0.4nm). The spreading tends to be less severe near the $\Delta_V$ minimum, so the magnetic field can be chosen to operate close to this point. As $\Delta_V=g.\mu_B.B$ defines the anticrossing, traveling up the curve leads to the spin regime, and down to the valley regime. Separate back-Gates for each qubit would enable to adjust individually the $V_b$ range to toggle between the spin and valley regimes.